# Topology Optimization Design of Stretchable Metamaterials with Bezier Skeleton Explicit Density (BSED) Representation Algorithm

Hao Deng, Shawn Hinnebusch, and Albert C. To[*]

Department of Mechanical Engineering and Materials Science, University of Pittsburgh, Pittsburgh, PA 15261

*Corresponding author. Email: albertto@pitt.edu

**Abstract**

A new density field representation technique called the Bezier skeleton explicit density (BSED) representation scheme for topology optimization of stretchable metamaterials under finite deformation is proposed for the first time. The proposed approach overcomes a key deficiency in existing density-based optimization methods that typically yield designs that do not have smooth surfaces but have large number of small intricate features, which are difficult to manufacture even by additive manufacturing. In the proposed approach, Bezier curves are utilized to describe the skeleton of the design being optimized where the description of the entire design is realized by assigning thickness along the curves. This geometric representation technique ensures that the optimized design is smooth and concise and can easily be tuned to be manufacturable by additive manufacturing. In the optimization method, the density field is described by the Heaviside function defined on the Bezier curves. Compared to NURBS or B-spline based models, Bezier curves have fewer control parameters and hence are easier to manipulate for sensitivity derivation, especially for distance sensitivities. Due to its powerful curve fitting ability, using Bezier curve to represent density field allows exploring design space effectively and generating concise structures without any intricate small features at the borders. Furthermore, this density representation method is mesh independent and design variables are reduced significantly so that optimization problem can be solved efficiently using small-scale optimization algorithms such as sequential quadratic programming. Numerical optimization results under three typical tests, uniaxial tension, biaxial tension and shear tests are presented by imposing symmetry on a unit cell for lattice structure design. Numerical examples illustrate that the optimized metamaterials have better stretchability and higher stiffness, and hence the proposed method has great potential of impacting the design of future applications that exploiti stretchability of structures.

Keywords: Topology optimization, Stretchable metamaterial, Bezier-based, SIMP

## 1. Introduction

Stretchable electronics has been studied for almost 20 years and several novel applications ranging from bio-integrated devices to wearable technologies can be found. Demand for higher performing mechanical design raises new challenges in soft system designs. For instance, wearable electronics, which deals with complex, flexible and stretchable biological systems, require that artificial material to be able to exhibit high stretchability while retaining stiff to transfer loading. Unfortunately, conventional electronics made of silicon or polymers are rigid and brittle in nature and hence are not ideal for wearable electronics due to lacking the ability to stretch. A number of ground-breaking ideas have recently been proposed to achieve the above functional requirements such as (I) application of unconventional materials (e.g. hydrogel [1]) and (II) novel structures with new mechanical characteristics (e.g. serpentine-shaped structures [2]). In this paper, we focus on designing new metamaterial to achieve certain functionality such as stretchability and compliance. With the advent of additive manufacturing technology, the ability to fabricate complicated geometries made of varies materials from metal to soft materials is possible. Thus, metamaterial design becomes an emerging field in research in that it may be utilized to generate novel material to satisfy the desired functional requirement. As reported in [3], some designed lattice structures made of metamaterials show ultrahigh reversible stretchability, which opens the door to design stretchable electronics.

Topology optimization is an advanced computational method that automatically optimizes material layout within a given design space given certain functional and weight requirements. Over the last two decades, topology optimization presents powerful applicability to create novel materials with enhanced properties, such as materials with negative thermal expansion coefficient [4] or photonic materials [5, 6]. Several other remarkable works are achieved by researchers regarding metamaterial design. F. Wang [7] designed materials based on topology optimization to achieve prescribed nonlinear properties using finite deformation theory. Andreassen [8] developed an optimization algorithm to design manufacturable extremal elastic materials, which possess unique properties such as negative Possion's ratio. Sigmund [9] created a new class of architected materials using topology optimization with programmable Possion's ratio with 3D printing to fabricate the designs. Recently, Wang [10] presented a topology optimization framework for designing periodic cellular materials with maximized strength under compressive loading, where failure mechanism of buckling instability is considered.

Compared to traditional metamaterial design optimization performed using linear finite element analysis, designing metamaterial with high stretchability requires considering geometrical and material nonlinearity. Determining the effective homogenized properties of nonlinear materials at finite deformation is challenging and is an active field of research [11, 12]. Elastomer test [13], such as uniaxial tension, is one feasible method to determine effective properties of novel metamaterial. Similar to the methods applied in Ref. [14], three elastomer testing methods (uniaxial tension, equi-biaxial tension, pure shear) are applied in this paper to evaluate effective material properties.

Although topology optimization method for material design has already achieved remarkable achievements in recent years, there still exist several challenges such as controlling the structural complexity and ensuring manufacturability of an optimal design. To ensure manufacturability of the optimized design, different methods have been reported for controlling geometric complexity and minimum length in design. Bendsoe and co-workers [15] proposed a well-known ground structure approach in an explicit way to control structural complexity with a predefined initial structures. Based on the ground structure method, several researchers [16-18] further developed this method to accommodate various situations. To control design complexity in an explicit geometrical way, a Moving Morphable Component (MMC) approach is proposed by Guo et al [19]. In their method, all components are described by the level set functions and allowed to move, overlap, and merge freely, where the Extended Finite Element Method (XFEM) based on a fixed mesh is employed to solve the physical problem. Based on the MMC approach, Guo et al [20-22] further extended the MMC to resolve 3-D problem and improved this method to solve more complex physical problems such as stress constraint optimization and multiple materials design. Recently, Tortorelli and co-workers [23] proposed a continuum-based topology optimization method to design discrete structural elements called the geometric projection method. Tortorelli's method is based on the density-based topology optimization framework, and hence standard finite element progress and nonlinear programming algorithm can be applied, where a differentiable mapping from discrete element to density field is realized in this paper. Furthermore, Narato et al [24] further developed this method to solve the constrained stress problem, where the optimal design is made by assembling discrete geometric components like bars or plates. Lately, Tortorelli [25] extended their geometric projection work to three dimensions and design unit cell for lattice materials based on inverse homogenization, where a negative Poisson's ratio lattice material is achieved. Recently, Daniel et al [26] proposed a novel method to represent the density field with a truncated Fourier representation, where the number of decision variables are reduced significantly. In fact, geometry projection methods and Fourier representation can be classified as a dimension reduction method. From the mathematical point of view, the key problem is to find an appropriate density field representation methodology to take the place of traditional density-based method. Traditional density-based method, where each element works as a design variable, always results in complex geometry with large number of small intricate features, while these small features are not amenable for manufacturability for AM or post-processing that can cause a loss in geometric accuracy.

To address the above challenge, a new density field representation technique called the Bezier skeleton explicit density (BSED) representation scheme for topology optimization of stretchable metamaterials is proposed in this paper for the first time. First, Bezier curves are widely used in computer graphics to produce curves which appear reasonably smooth at all scales and are employed in the proposed approach to describe the skeleton of the design being optimized so that the entire design is described by assigning the Bezier curve with certain thickness. The proposed approach ensures that the optimized design is smooth and concise, and can easily be manufactured by AM. Second, the density field is described by the Heaviside function defined on the Bezier curves in the optimization model. Compared to NURBS or B-spline based models, Bezier curves have fewer control parameters and hence are easier to manipulate for sensitivity derivation, especially for distance sensitivities. Due to its powerful curve fitting ability, using Bezier curve to represent density field allows exploring design space effectively and generating concise structures without any intricate small features at the borders. Furthermore, this density representation method is mesh independent and design variables are reduced significantly so that optimization problem can be solved efficiently using small-scale optimization algorithms such as sequential quadratic programming.

The paper is organized as follows. In Section 2, nonlinear finite element formulation is introduced. Section 3 describes the Bezier-based explicit density representation algorithm in details. In Section 4, the physical models are demonstrated to characterize the material properties in standard elastomer tests under finite deformation. Section 5 describes the failure phenomenon for soft material. In Section 6, the optimization problem is formulated for stretchable metamaterial design. Section 7 presented several optimized numerical results for stretchable material design. The paper ends with some conclusions in Section 8.

## 2. Nonlinear Finite Element Formulation

The analysis model of a hyperelastic body under external loading is briefly described in this section. A point locates at $X$ at initial state transfers to $x$ in the deformed configuration, and the displacement vector is $u = x - X$. Transformation can be described by deformation gradient $F$,

$$F = I + \frac{\partial u}{\partial X} = I + \nabla_0 u \tag{1}$$

Note that $I$ is second-order identity tensor. Spatial equilibrium equation for a deformable body is written as,

$$div\ \sigma + f = 0 \tag{2}$$

where $\sigma$ is Cauchy stress tensor and $f$ is body force. In this paper, the Mooney-Rivlin model [27], which is one of the most popular hyperelastic material model, is adopted here to describe the strain energy function.

To simplify our problem, conservative external force is assumed here, which is independent of structural deformation. Using a finite element discretization, the structural equilibrium can be written as

$$\boldsymbol{r} = \boldsymbol{f}^{ext} - \boldsymbol{f}^{int} \tag{3}$$

where $\boldsymbol{f}^{ext}$ is the external nodal load vector, $\boldsymbol{f}^{int}$ is the internal nodal load vector, and $\boldsymbol{r}$ is residual vector [13]. It is necessary to note that both material nonlinearity and geometric nonlinearity need to be taken into consideration during hyperelastic material deformation analysis. The three invariants of the right Cauchy-Green deformation tensor are expressed as follows:

$$\begin{aligned} I_1 &= \boldsymbol{C} : \boldsymbol{I} \\ I_2 &= \boldsymbol{C} : \boldsymbol{C} \\ I_3 &= det(\boldsymbol{C}) \end{aligned} \tag{4}$$

where $\boldsymbol{C}$ is the right Cauchy–Green deformation tensor, and $\boldsymbol{I}$ is identity tensor. Mathematical operator (:) denotes double contraction of two tensors. The strain energy expression of Mooney-Rivlin model, which includes the effect of the first and second invariants $I_1$ and $I_2$, can be written as follows:

$$\psi(J_1, J_2, J_3) = A_{10}\left(I_1 I_3^{-1/3} - 3\right) + A_{01}\left(I_2 I_3^{-2/3} - 3\right) + \frac{K}{2}(I_3^{1/2} - 1)^2 \tag{5}$$

where $A_{10}$ and $A_{01}$ are two nonzero parameters, which need to be determined through material testing. $K$ is the bulk modulus. Most hyperelastic materials such as rubber have a large bulk modulus, which means a small volume change leads to a large hydrostatic pressure.

## 3. Bezier-based Explicit Density Representation Algorithm

To control the structural complexity and minimum length of an optimal design, a new density representation algorithm is proposed in this section to explicitly describe and control geometric shape using Bezier-based skeleton. Compared to the bar-based density representation method proposed by Norato et al [23, 28] and Tortorelli et al [25], Bezier-based explicit density representation algorithm has more freedom to explore the design space and generate concise curved structures.

### 3.1 Bezier representation

To represent the skeleton based on curve segment, the so-called Bezier polygon is introduced. Every polynomial curve segment can be represented by Bezier polygon, where the curve segment lies in the convex hull of its Bezier polygon. Note that a Bezier curve is a parametric curve that uses Bernstein polynomials as a basis.

### 3.1.1 Bernstein polynomials

The Bernstein polynomials of degree $n$ is defined as follows,

$$B_i^n(u) = \binom{i}{n} u^i (1-u)^{n-i}, i = 0, \cdots, n \tag{9}$$

where

$$\binom{i}{n} = \frac{n!}{i!(n-i)!} \tag{10}$$

The Bernstein polynomials over [0,1] of different degree $n$ (4, 5, 6) are plotted in Fig. 1.

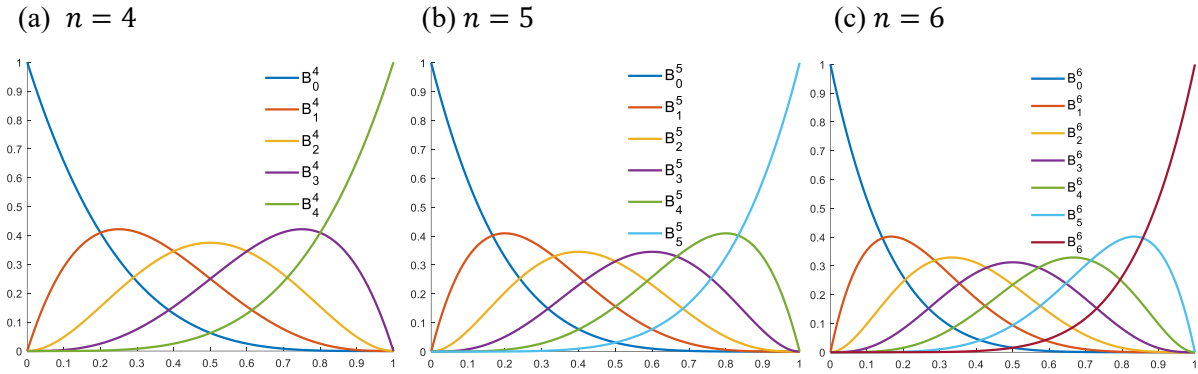

Figure 1. The Bernstein polynomials of degree $n$ over [0, 1]: (a) $n$=4, (b) $n$=5, and (c) $n$=6.

All polynomials of degree less than $n$ can be represented by Bernstein polynomials bases $B_i^n$, and hence the Bezier representation of polynomial curve $\chi(u)$ is defined as:

$$\chi(u) = \sum_{i=0}^{n} \psi_i B_i^n(u) \tag{11}$$

where $\psi_i$ is coefficient of the Bernstein polynomial bases. Using the following affine transformation

$$u = a(1-t) + bt, \quad a \neq b, t \in [0,1], \tag{12}$$

the $n$ th degree Bezier representation of polynomial curve can be written as:

$$\chi(t) = \sum_{i=0}^{n} \chi_i B_i^n(t) \tag{13}$$

where coefficients $\chi_i$ are called Bezier points, which are the vertices of the Bezier polygon of curve $\chi(t)$ over the interval $[a, b]$. Parameter $t$ is called local parameter. A high-order Bezier curve segment with five control points is shown in Fig. 2, where the blue dotted line and blue dots represent Bezier polygon and control points, respectively. Equation (13) can be expressed explicitly as follows:

$$\chi(t) = \sum_{i=0}^{n} \binom{n}{i} (1-t)^{n-i} t^i \chi_i \quad (t \in [0,1]) \tag{14}$$

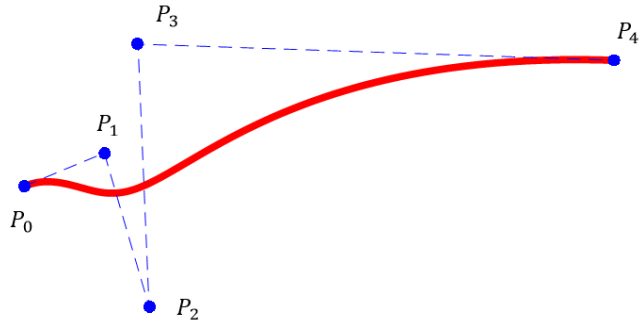

Figure 2. A high-order ($n$=4) curve segment with its Bezier polygon

## 3.2 Geometry mapping based on Heaviside function

The design is defined by a set of curved-based skeletons, which control the density distribution using the Heaviside function. Each curved-based skeleton is described by a single Bezier curve. The width of the skeleton is determined by a parameter $w$ in the Heaviside function. The minimum distance from any point $\boldsymbol{p}$ in the design domain to the skeleton curve is demonstrated in Fig. 3, where the blue solid line represents the minimum distance. Given a point $\boldsymbol{p}$ and a Bezier curve $\chi(t)$, the point projection (minimum distance) can be described as finding a solution $t^*$, such that,

$$\|\boldsymbol{p} - \chi(t^*)\| = min\{\|\boldsymbol{p} - \chi(t)\|\}, \ t \in [0,1] \qquad (6)$$

If $t^* \notin \{0.1\}$, the following necessary condition should be satisfied,

$$\chi'(t^*) = 0 \qquad (7)$$

where $\chi'(t)$ denotes the derivative of $\chi(t)$. Thus, the above point projection problem can be solved by a root-finding problem of a polynomial equation, and more implementation details can be found in Ref [29].

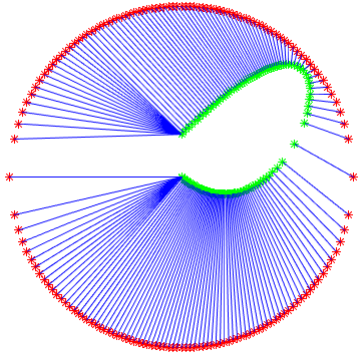

Figure 3. Point projection on a curved skeleton

To perform topology optimization algorithm on a fixed grid, geometric mapping from curved-based skeletons to a density field is achieved by a smoothed Heaviside function stated as:

$$\rho = \frac{1}{2}\bigl(1 + \tanh(\beta(w - d))\bigr)\bar{\bar{\rho}} + \rho_{min} \qquad (8)$$

where $\bar{\bar{\rho}}$ denotes density of the segment, and the segment can be considered as non-existent if $\bar{\bar{\rho}} = 0$. $d$ represents the projection distance from centroid of grid to skeleton. $w$ is a threshold used to determine the width of mapping domain and parameter $\beta$ determines the properties of density transition region. The parameter $\beta$ has significant effect on the boundary of geometric projection as shown in Fig. 4. Increasing value of $\beta$ makes boundary become more distinct, and width of geometry mapping is determined by $w$ as plotted in Fig. 5. $\rho_{min}$ is a small non-negative value. Obviously, the thickness of skeleton can be modified directly through parameter $w$, which is able to control the minimum length of the optimized design.

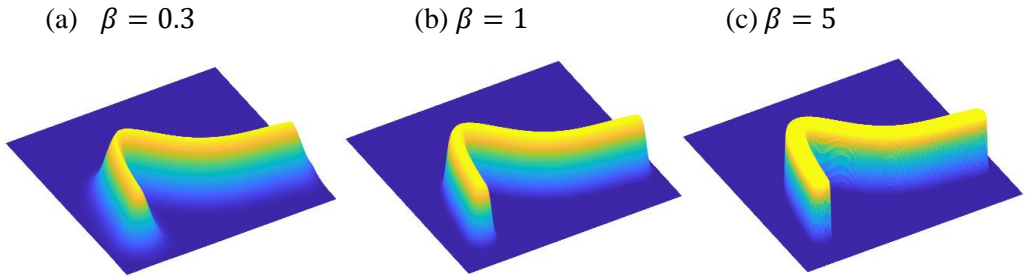

Figure 4. The effect of parameter $\beta$ on density field (a) $\beta = 0.3$ (b) $\beta = 1$ (c) $\beta = 5$

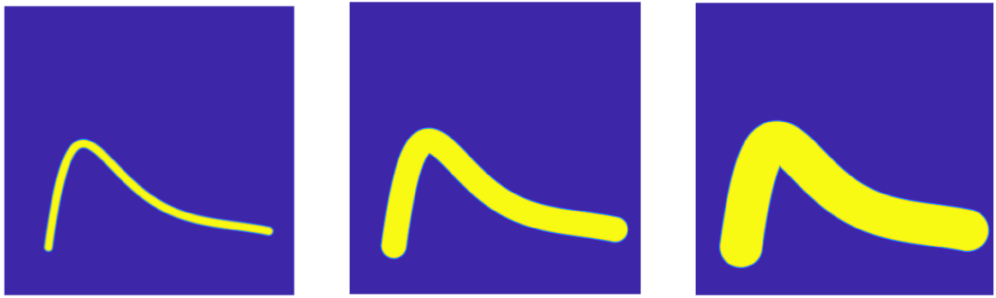

Figure 5. Effect of parameter $w$ on geometry mapping: (a) $w = 1$, (b) $w = 3$, and (c) $w = 5$ for $\beta = 5$

To explicitly express geometry control based on curved skeleton, the relationship between curved skeleton and mapping density field is plotted in Fig. 6. Figure 6(c) plotted the boundary enveloping line of mapping density, which can effectively reflect the effect of skeleton thickness on density field. Note that the boundary enveloping line becomes non-smooth if mesh density of FEM is not enough.

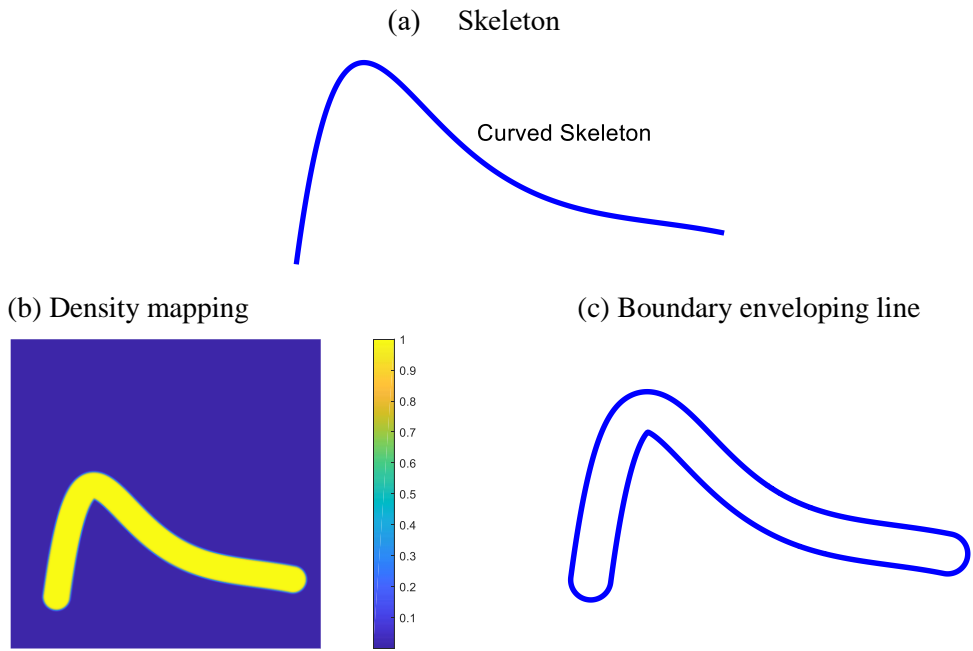

Figure 6. Geometry mapping from skeleton to density field (a) skeleton, (b) density mapping, and (c) boundary enveloping line

## 3.3 Density field mapping of multiple curved components

The previous section describes the mapping from a single component to density layout. For multiple curved components, composite density needs to be defined as follows:

$$\widetilde{\rho}_j = \max \rho_{ij} \ (i = 1,2 \cdots n, j = 1,2 \cdots m) \tag{15}$$

where $n$ denotes number of components and $m$ represents total element number. Due to the non-differentiable nature of maximum function, p-norm formulation is applied to achieve smooth approximation of the maximum function. Thus, the composite density is defined as:

$$\widetilde{\rho}_j = \left(\sum_{i=1}^{n} \rho_{ij}^{p}\right)^{1/p} \tag{16}$$

Note that if $p$ tends to $+\infty$, the value in the p-norm formulation above approximates the maximum of density $\rho_{ij}$, while for finite $p$ value, p-norm function always exceeds the maximum density. In this paper, the value of $p$ is set to be 10. As mentioned in Ref. [25], composite density may exceed unity. However, for two-dimension design, it is necessary to restrict composite density between 0 and 1. To overcome this numerical issue, a special density projection (SDP) function is introduced as follows:

$$\bar{\rho}_j = \tanh(3\widetilde{\rho}_j) \tag{17}$$

The curve property of the above projection function is shown in Fig. 7. Compared to Ref. [25], differentiability during optimization is guaranteed using SDP function instead of applying discontinuous minimum function to avoid artificially high stiffness ($\widetilde{\rho}_j > 1$) at local regions. Note that $\bar{\rho}_j$, which is the actual input data for FEM analysis, represents the real physical density in this work.

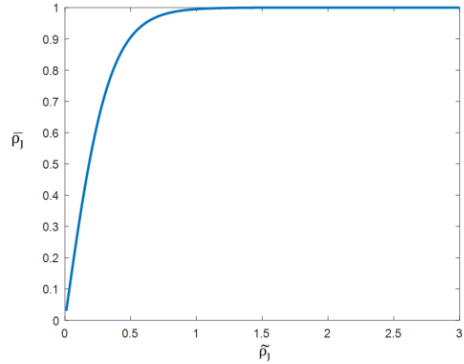

Figure 7. Properties of special density projection function

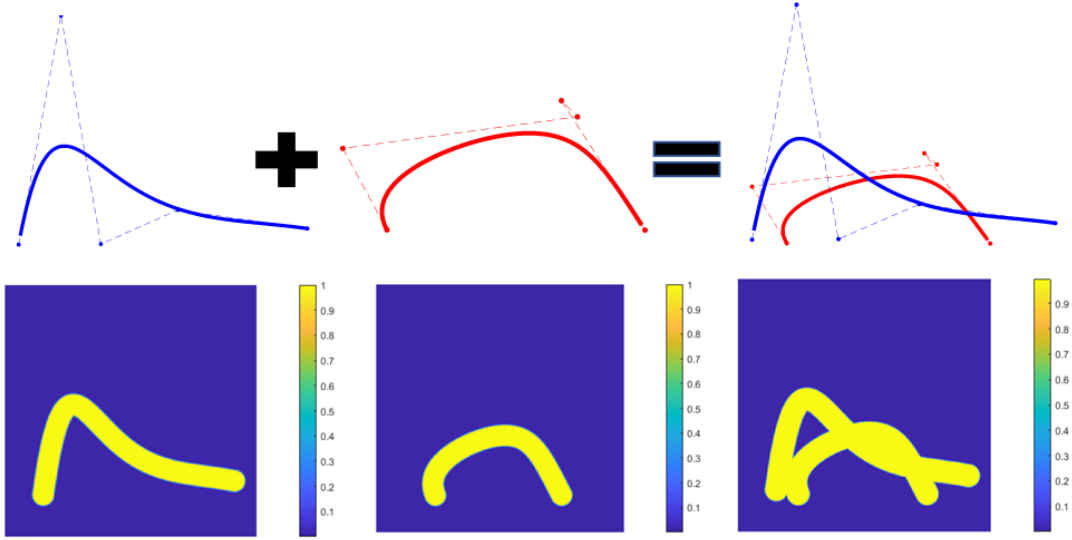

Figure 8. Density field mapping of two curved components ($w = 3, \beta = 5$)

### 3.4 Sensitivity analysis of Bezier-based explicit density representation

To solve the topology optimization problem using Bezier-based explicit density representation method, the sensitivity of the density field with respect to geometry control parameters **X** is required. The sensitivity of composite density $\bar{\rho}_J$ with respect to control parameters **X** can be obtained by the chain rule as following,

$$\frac{\partial \bar{\rho}_J}{\partial \mathbf{X}} = \frac{\partial \bar{\rho}_J}{\partial \tilde{\rho}_J} \frac{\partial \tilde{\rho}_J}{\partial \rho_{ij}} \frac{\partial \rho_{ij}}{\partial \mathbf{X}} \tag{18}$$

where $\mathbf{X} = [\chi_i, \bar{\rho}, w]$ $(i = 1,2,\cdots,n)$. $\chi_i$ represents control points of Bezier curves. The first two terms in above equations are given by,

$$\frac{\partial \bar{\rho}_J}{\partial \tilde{\rho}_J} = 3 - 3\tanh(3\tilde{\rho}_J)^2 \tag{19}$$

$$\frac{\partial \tilde{\rho}_J}{\partial \rho_{ij}} = \left(\sum_{i=1}^{n} \rho_{ij}^{p}\right)^{\frac{1}{p}-1} \left(\sum_{i=1}^{n} \rho_{ij}^{p-1}\right) \tag{20}$$

The derivatives of $\rho_{ij}$ with respect to projection parameters $\bar{\bar{\rho}}_k$ and $w_k$ are given as following,

$$\frac{\partial \rho_{ij}}{\partial \bar{\bar{\rho}}_k} = \frac{1}{2}(1 + \tanh(\beta(w - d_{ij}))\delta_{ik} \tag{21}$$

$$\frac{\partial \rho_{ij}}{\partial w_k} = \frac{1}{2}\bar{\bar{\rho}}_i(\beta(1 - \tanh\left(\beta(w_k - d_{ij})\right)^2) \tag{22}$$

k denotes the index of geometry component. where $\delta_{ik}$ is Kronecker delta, defined as:

$$\begin{cases} \delta_{ik} = 1 \; if \; i = k \\ \delta_{ik} = 0 \; if \; i \neq k \end{cases} \tag{23}$$

$\frac{\partial \rho_{ij}}{\partial \chi_i}$ can be obtained by chain rule as follows,

$$\frac{\partial \rho_{ij}}{\partial \chi_i} = \frac{\partial \rho_{ij}}{\partial d_{ij}} \frac{\partial d_{ij}}{\partial \chi_i} \tag{24}$$

To obtain the derivative of $d_{ij}$ with respect to the Bezier points $\chi_i = [a_i, b_i]$, we rewritten the curve $\chi(t) = [x(t), y(t)]$ in an explicit way as follows,

$$x(t) = \sum_{i=0}^{N} \binom{n}{i}(1-t)^{n-i} t^i a_i \tag{25}$$

$$y(t) = \sum_{i=0}^{N} \binom{n}{i}(1-t)^{n-i} t^i b_i \tag{26}$$

Note that $d_{ij}$ represents minimum distance from centroid of element $j$ to curved component. Minimum distance $d$ from arbitrary point $(x_0, y_0)$ to a given Bezier curve can be written as

$$d = \sqrt{(x(t_0) - x_0)^2 + (y(t_0) - y_0)^2} \tag{27}$$

where $(x(t_0), y(t_0))$ is the projection point on Bezier curve, and local coordinate $t_0$ should satisfy

$$\chi'(t_0) = 0 \ (if \ t_0 \notin \{0,1\}) \tag{28}$$

The derivatives of minimum distance $d$ with respect to the Bezier points $\chi_i = (a_i, b_i)$ are,

$$\frac{\partial d}{\partial a_i} = [(x(t_0)-x_0)^2 + (y(t_0)-y_0)^2]^{-\frac{1}{2}} \left[(x(t_0)-x_0)\frac{\partial x(t_0)}{\partial a_i} + (y(t_0)-y_0)\frac{\partial y(t_0)}{\partial a_i}\right] \tag{29}$$

$$\frac{\partial d}{\partial b_i} = [(x(t_0)-x_0)^2 + (y(t_0)-y_0)^2]^{-\frac{1}{2}} \left[(x(t_0)-x_0)\frac{\partial x(t_0)}{\partial b_i} + (y(t_0)-y_0)\frac{\partial y(t_0)}{\partial b_i}\right] \tag{30}$$

Note that

$$\begin{aligned} \frac{\partial x(t_0)}{\partial a_i} &= \frac{\partial x(t_0)}{\partial a_i} + \frac{\partial x(t_0)}{\partial t_0}\frac{\partial t_0}{\partial a_i} & \frac{\partial y(t_0)}{\partial a_i} &= \frac{\partial y(t_0)}{\partial t_0}\frac{\partial t_0}{\partial a_i} \\ \frac{\partial x(t_0)}{\partial b_i} &= \frac{\partial x(t_0)}{\partial t_0}\frac{\partial t_0}{\partial b_i} & \frac{\partial y(t_0)}{\partial b_i} &= \frac{\partial y(t_0)}{\partial b_i} + \frac{\partial y(t_0)}{\partial t_0}\frac{\partial t_0}{\partial b_i} \end{aligned} \tag{31}$$

where the term $\frac{\partial x(t_0)}{\partial a_i}$ and $\frac{\partial y(t_0)}{\partial b_i}$ can be easily obtained as follows,

$$\frac{\partial x(t_0)}{\partial a_i} = \frac{\partial y(t_0)}{\partial b_i} = \binom{n}{i}(1-t_0)^{n-i} t_0^i \tag{32}$$

$t_0$ needs to satisfy

$$(x(t_0) - x_0)\frac{\partial x(t_0)}{\partial t_0} + (y(t_0) - y_0)\frac{\partial y(t_0)}{\partial t_0} = 0 \tag{33}$$

Simultaneous derivation on both sides of the equation with respect to Bezier parameter $a_i$,

$$\left(\frac{\partial x(t_0)}{\partial a_i} + \frac{\partial x(t_0)}{\partial t_0}\frac{\partial t_0}{\partial a_i}\right)\frac{\partial x(t_0)}{\partial t_0} + (x(t_0) - x_0)\left[\frac{\partial Dx(t_0)}{\partial a_i} + \frac{\partial Dx(t_0)}{\partial t_0}\frac{\partial t_0}{\partial a_i}\right] + \frac{\partial y(t_0)}{\partial t_0}\frac{\partial t_0}{\partial a_i}\frac{\partial y(t_0)}{\partial t_0} + (y(t_0) - y_0)\frac{\partial Dy(t_0)}{\partial t_0}\frac{\partial t_0}{\partial a_i} = 0 \tag{34}$$

where

$$Dx(t_0) = \frac{\partial x(t_0)}{\partial t_0} = n\sum_{i=0}^{n-1}\binom{n-1}{i}(1-t_0)^{n-1-i} t_0^i c_i \tag{35}$$

$$Dy(t_0) = \frac{\partial y(t_0)}{\partial t_0} = n\sum_{i=0}^{n-1}\binom{n-1}{i}(1-t_0)^{n-1-i} t_0^i d_i \tag{36}$$

$$\frac{\partial Dx(t_0)}{\partial t_0} = n(n-1)\sum_{i=0}^{n-2}\binom{n-2}{i}(1-t_0)^{n-2-i} t_0^i (c_{i+1} - c_i) \tag{37}$$

$$\frac{\partial Dy(t_0)}{\partial t_0} = n(n-1) \sum_{i=0}^{n-2} \binom{n-2}{i} (1-t_0)^{n-2-i} t_0^{i} (d_{i+1} - d_i) \tag{38}$$

Eq. (34) can be simplified as following,

$$\frac{\partial t_0}{\partial a_i} = -\frac{Q_a}{P_a} \tag{39}$$

$$P_a = \left(\frac{\partial x(t_0)}{\partial t_0}\right)^2 + (x(t_0) - x_0)\frac{\partial Dx(t_0)}{\partial t_0} + \left(\frac{\partial y(t_0)}{\partial t_0}\right)^2 + (y(t_0) - y_0)\frac{\partial Dy(t_0)}{\partial t_0} \tag{40}$$

$$Q_a = \frac{\partial x(t_0)}{\partial a_i}\frac{\partial x(t_0)}{\partial t_0} + (x(t_0) - x_0)\frac{\partial Dx(t_0)}{\partial a_i} \tag{41}$$

Similarly, $\frac{\partial t_0}{\partial b_i}$ can be obtained as follows,

$$\frac{\partial t_0}{\partial b_i} = -\frac{Q_b}{P_b} \tag{42}$$

$$P_b = \left(\frac{\partial y(t_0)}{\partial t_0}\right)^2 + (y(t_0) - y_0)\frac{\partial Dy(t_0)}{\partial t_0} + \left(\frac{\partial x(t_0)}{\partial t_0}\right)^2 + (x(t_0) - x_0)\frac{\partial Dx(t_0)}{\partial t_0} \tag{43}$$

$$Q_b = \frac{\partial y(t_0)}{\partial b_i}\frac{\partial y(t_0)}{\partial t_0} + (y(t_0) - y_0)\frac{\partial Dy(t_0)}{\partial b_i} \tag{44}$$

Substituting Eq. (42), Eq. (39), and Eq. (31) into Eq. (29) and Eq. (30), the derivatives of minimum distance $d$ with respect to the Bezier points $\chi_i = [a_i, b_i]$ can be obtained. Note that for $t_0 \in \{0,1\}$, the sensitivities are easily obtained.

## 4. Characterization of Material Behavior in Elastomer Test

For a periodic material, effective material properties can be evaluated using a unit cell as illustrated in Fig. 9. Homogenization methods can be regarded as an effective method to calculate material properties for small deformation problems. For finite deformation problem, the effective material properties strongly depends on each deformation state and accurately predict material nonlinear behavior is still a challenging problem [30, 31]. Applying standard elastomer testing method is one feasible way to simplify the problem. Three major strain states including uniaxial tension, equal biaxial tension, and simple shear are applied in this paper to evaluate effective material properties, and designs under three different strain states are presented to exhibit high stretchability and stiffness.

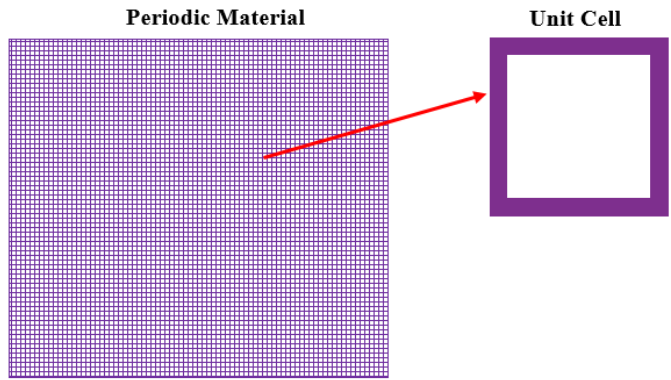

Figure 9. Schematic illustration of a unit cell and a periodic material

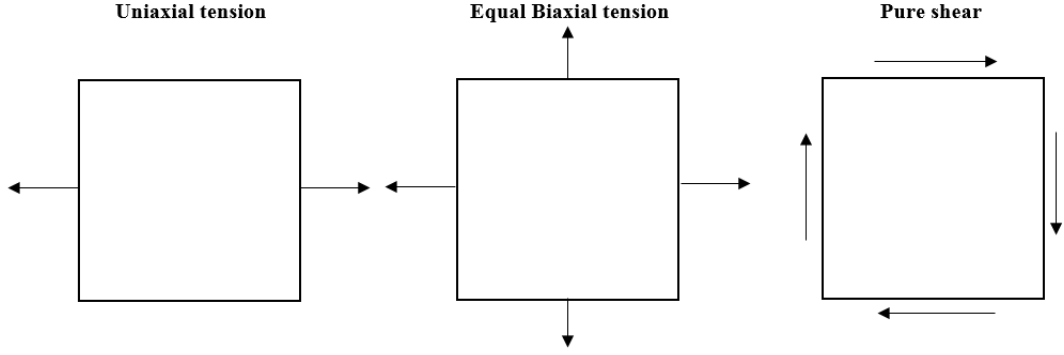

Figure 10. Three major strain states

In the elastomer tests, the material behavior can be characterized using the unit cell under periodic boundary conditions. For uniaxial tension test, materials are uniformly stretched along longitudinally as shown in Fig. 10. In such a situation, each node on the left and right boundaries need to satisfy the periodic boundary condition, which means that a constant displacement difference $u_c$ is assumed such that $u_r - u_l = u_c$. Note that $u_r$ and $u_l$ represent displacement on right and left boundary, respectively. It is important to mention that the transverse displacement for upper and lower boundary should also satisfy periodic boundary conditions. More details regarding applying periodic boundary conditions can be found in Ref. [32].

### 4.1 Periodic boundary conditions

For finite element analysis, multi-point constraints (MPCs) is an effective way to impose periodic boundary conditions [33]. MPCs enforce relations among the degrees of freedoms at two or more distinct nodes in a FE model. For periodic boundary conditions, a set of linear equations that couple the DOFs by the constraints, are called "constraint equations". In such a situation, the constraint equations can be written as,

$$\boldsymbol{Au = Q} \tag{45}$$

where $\boldsymbol{A}$ is a constant matrix and $\boldsymbol{Q}$ is a constant vector. $\boldsymbol{u}$ is global displacement vector. The schematic illustration of periodic boundary conditions is demonstrated in Fig. 11.

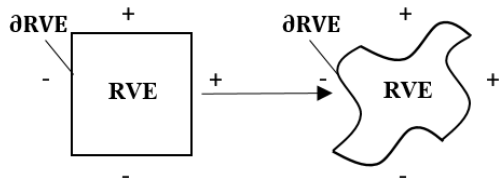

Figure 11. Schematic illustration of periodic boundary conditions

To implement the above equations in nonlinear finite element analysis. Three different methods are available to impose above constraint equations, namely: (1) Transformation equation method, (2) Lagrange multiplier method, and (3) penalty function method. In this paper, the penalty function method is applied to impose periodic boundary conditions in FEM analysis. For nonlinear finite element with constraint equations, the following equilibrium equation is constructed [33]:

$$\boldsymbol{R(u)} + (\boldsymbol{A^T \alpha A})\boldsymbol{u} - \boldsymbol{F} - \boldsymbol{A^T \alpha Q} = \boldsymbol{0} \tag{46}$$

where $R(u)$ is internal force and $F$ is external force. $\alpha$ is diagonal matrix of penalty weights, with $\alpha_{ii} > 0$ and $\alpha_{ij} = 0$ $(i \neq j)$. It is worth to mention that penalty term in above equations can be physically interpreted as additional forces to enforce the constraint approximately. Note that $A^T \alpha A$ is referred to as penalty matrix, and the constraints can be satisfied exactly if the penalty weight $\alpha_{ii}$ tends to infinity. Actually, choosing a right penalty need to balance the trade-off between reducing the constraint violation and limiting the solution error due to ill-conditioning system. In this paper, the value of penalty weight is chosen as $\alpha_{ii} = 10^8$.

## 5. Generalized Energy Failure Criterion for Hyperelastic Materials

When designing stretchable metamaterial, the critical problem is that how to guarantee material reversible capacity under large deformation without plastic deformation or fracture. For regular 3D printing materials, the elastic regime of these materials is usually limited to 10% or less, such as ABS (Acrylonitrile butadiene styrene). For inorganic electronic materials such as silicon, small strains (around 1%) can lead to rupture. Therefore, the primary goal of designing stretchable metamaterial is resisting irreversible deformation. For metallic materials, von Mises stress or strain can work as a criterion to measure material failure behavior, while the failure mechanism is more sophisticated for soft materials. For centuries, scientists have made great efforts to develop theories for predicting mechanical failure of materials, and eventually the "generalized energy criterion" is proposed to be one universal law for various different kinds of materials. Based on continuum thermodynamics, material failure in solid material is triggered by internal interactions between material particles, which is represented macroscopically by a specified elastic strain energy density as described in Ref. [34]. Energy-based failure criterion is regarded as a universal criterion for different types of materials. According to Ref. [35], the failure of materials usually originates from a specific plane, and basic failure mechanisms contain two physical conditions: Shearing failure driven by shear stress and cleavage driven by the normal stress as presented in Fig. 12. Fracture of material needs energy to break the atomic bonding to form the crack, and the energy density associated with the failure is defined presented in Fig. 12.

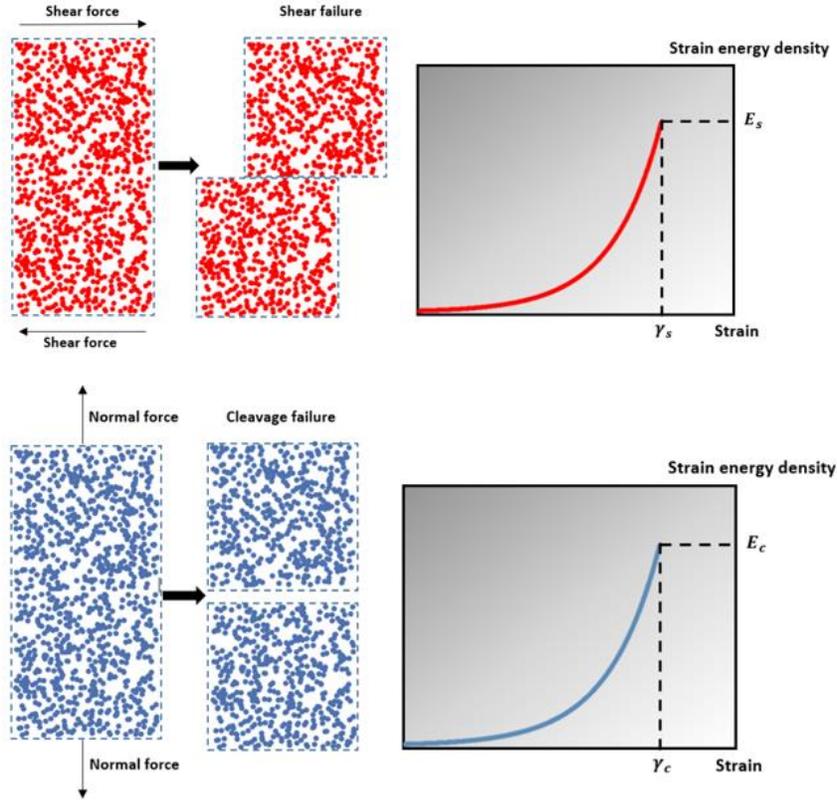

Figure 12. Illustrations on the two basic failure mechanisms of material

When designing soft materials, material models such as the Mooney-Rivlin model cannot model the softening of the material at large strains since the model assumes that the strain energy of the model materials can increase without bound. However, it is clear that no real material can store unlimited amount of energy without failure under finite deformation. As described in Ref. [36], real hyperelastic materials may experience softening when material strain is large enough. Volokh [37-41] further developed this theory and proposed the softening hyperelasticity approach, which describes strain softening by introducing the energy limiter. Meanwhile, Rittel et al [42] have observed the existence of material energy limiter in experiments. Traditional continuum-based hyperelastic models do not include energy limiter which leads to unbounded energy accumulation. Evidently, this is unphysical and may result in unreasonable engineering design when using traditional hyperelastic materials. To design stretchable metamaterial, the energy of real hyperelastic material needs to be limited the design load and may be defined as material failure energy. Such a limiter is a direct criterion to measure recoverability of material. In fact, different failure criteria have been utilized to describe failure of hyperelastic materials, which include the (1) maximum principal stress, (2) maximum principal stretch, (3) maximum shear stress, (4) von Mises stress, and (5) strain energy. Based on the experiment conducted by Volokh [43], the results show that strain energy is almost constant for the failure states induced by various loading modes ranging from uniaxial to equal biaxial tension. The von Mises stress exhibits a wider range of scattering as compared to strain energy. The maximum stresses and stretches vary significantly with the variation of loads from uniaxial to equal biaxial tension. Thus, using strain energy as failure criterion is more accurate and reasonable for measuring failure of soft material as described in Ref. [44] and hence will be utilized also in the optimization algorithm development.

## 6. Optimization Method

This study focuses on designing materials to present better stretchability and stiffness under finite deformation. The optimization formulation is described in detail in this section.

### 6.1 Design parameterizations

For structures experiencing large strains, excessive mesh distortion in low-density region often occurs, which always leads to divergence of nonlinear finite element analysis. Remeshing can alleviate this issue but is a cumbersome and computationally expensive process. To make optimization robust, an energy interpolation form proposed by Wang et al [45] is adopted here to relieve excessively distorted mesh in low-density area:

$$\Phi_e(\boldsymbol{u}_e) = [\Phi(\gamma_e \boldsymbol{u}_e) - \Phi_L(\gamma_e \boldsymbol{u}_e) + \Phi_L(\boldsymbol{u}_e)]E_e \quad (47)$$

where $\Phi()$ is the stored elastic energy density for base material and $\Phi_L()$ is the stored elastic energy density under small deformation. $E_e$ is a scaling parameter (i.e. $E_e = 1$ for solid material and $E_e = \varepsilon$ ($\varepsilon$ is a very small value) for void region). Linear element is chosen to describe material deformation behavior for low-density elements, which is insensitive to large deformation. In contrast, high-density elements need to be analyzed by non-linear analysis. In the equation above, the interpolation factor $\gamma_e$ equals to unity for solid elements, while $\gamma_e = 0$ corresponds to void element. The interpolation factor should satisfy that the stored energy corresponds to linear energy when $\gamma_e = 0$, while the elastic energy is simply depicted by the nonlinear energy term if $\gamma_e = 1$. A continuous and smooth method based on the Heaviside projection function is applied to ensure a smooth and differentiable transition between these two regions, which is successfully tested by the fictitious domain approach [45]. Therefore, the threshold parameter $\gamma_e$ can be modeled as follows:

$$\gamma_e = \frac{\tanh(\beta_1 x_0) + \tanh\left(\beta_1(\bar{x}_e{}^{pl} - x_0)\right)}{\tanh(\beta_1 x_0) + \tanh(\beta_1(1 - x_0))} \quad (48)$$

where $x_0$ is a threshold used to determine the element behavior. In most cases, $x_0 = 0.01$ and $\beta_1 = 500$ are reasonable values to separate these two distinct regions in the optimization progress. The scaled parameter $E_e$ for each element can be interpolated as:

$$E_e = \bar{x}_e(\boldsymbol{x})^{pl}(1 - \varepsilon) + \varepsilon \quad (49)$$

where $\varepsilon$ is a very small value (i.e. $\varepsilon = 10^{-5}$). $pl$ is penalization parameter and is set to a value of 3 in this work unless otherwise stated. Note that $\boldsymbol{x}$ is the design variable.

### 6.2 Optimization formulations

The optimization problem for designing stretchable and stiff metamaterial can be formulated to maximize material stiffness with local failure constraint for a given finite strain:

$$\begin{aligned} \max \boldsymbol{f} &= \boldsymbol{l}^T \boldsymbol{f}^{int}(\boldsymbol{u}) \\ \text{subject to} &\begin{cases} \frac{V(\boldsymbol{x})}{|\Omega|} - v_f^* \leq 0 \\ c\left[\frac{1}{N}\sum_{i=1}^{N}\left(\frac{E_i(\boldsymbol{x})}{\bar{E}}\right)^p\right]^{\frac{1}{p}} < 1 \end{cases} \end{aligned} \quad (50)$$

where $\boldsymbol{l}$ is a zero vector with unit entries at the degrees of freedom on the boundary, and $\boldsymbol{l}^T \boldsymbol{f}^{int}(\boldsymbol{u})$ represents reaction force on the boundary due to prescribed displacement. $E_i(\boldsymbol{x})$ is strain energy for every element and $\bar{E}$ is prescribed energy limiter for material failure. Note that the p-norm formulation is applied to measure local maximum element strain energy and $c$ is an adaptive parameter. The details of p-norm formulation and adaptive parameter can be found in Ref. [46]. $N$ denotes element number, and p-norm parameter is chosen as $p = 10$. $V(\boldsymbol{x})$ is the volume of the design, $|\Omega|$ is the total volume of the initial fixed design domain, and $v_f^*$ is the prescribed volume constraint.

### 6.3 Sensitivity analysis

### 6.3.1 General sensitivity derivation based on adjoint method

Gradient-based optimization method is employed here to solve the optimization problem above efficiently by deriving accurate sensitivities of the objective function and constraints. In the current study, gradients can be evaluated analytically using the adjoint method. For adjoint method, a general formulation corresponding to the nonlinear model is described in this section. The governing equilibrium equation in residual form can be written as follows:

$$\boldsymbol{\Psi} = \boldsymbol{R} - \lambda \cdot \boldsymbol{P} = \boldsymbol{0} \tag{51}$$

Any other constraint equations for the physical problem are expressed as follows:

$$\boldsymbol{\mathcal{H}} = \boldsymbol{0} \tag{52}$$

For a given function $F$, which can work as objective or constraint, an augmented Lagrangian function $G$ is formulated based on the adjoint method:

$$G = F + \boldsymbol{\psi}^T \boldsymbol{\Psi} + \boldsymbol{\kappa}^T \cdot \boldsymbol{\mathcal{H}} \tag{53}$$

where $\boldsymbol{\psi}$ and $\boldsymbol{\kappa}$ are Lagrange multiplier. For arbitrary vectors $\boldsymbol{\psi}$ and $\boldsymbol{\mathcal{H}}$, the equation $G = F$ can be established. Thus, achieving the derivative $\frac{\partial G}{\partial \rho}$ is equivalent to obtaining the sensitivities of the augmented Lagrangian function $G$ with respect to material density $\rho$. A general procedure to obtain the derivative of $G$ using the chain rule can be expressed as:

$$\frac{\partial G}{\partial \rho} = \frac{\partial F}{\partial \rho} + \frac{\partial F}{\partial \boldsymbol{U}} \frac{\partial \boldsymbol{U}}{\partial \rho} + \boldsymbol{\psi}^T \left( \frac{\partial \boldsymbol{\Psi}}{\partial \rho} + \frac{\partial \boldsymbol{\Psi}}{\partial \boldsymbol{U}} \frac{\partial \boldsymbol{U}}{\partial \rho} \right) + \boldsymbol{\kappa}^T \cdot \left( \frac{\partial \boldsymbol{\mathcal{H}}}{\partial \rho} + \frac{\partial \boldsymbol{\mathcal{H}}}{\partial \boldsymbol{U}} \frac{\partial \boldsymbol{U}}{\partial \rho} \right) \tag{54}$$

where $\boldsymbol{U}$ is the global displacement vector and operator $\frac{\partial}{\partial \rho}$ represents a derivative with respect to $\rho$. Rearranging above Eq. (54) as follows:

$$\frac{\partial G}{\partial \rho} = \left( \boldsymbol{\psi}^T \frac{\partial \boldsymbol{\Psi}}{\partial \boldsymbol{U}} + \boldsymbol{\kappa}^T \frac{\partial \boldsymbol{\mathcal{H}}}{\partial \boldsymbol{U}} + \frac{\partial F}{\partial \boldsymbol{U}} \right) \frac{\partial \boldsymbol{U}}{\partial \rho} + \frac{\partial F}{\partial \rho} + \boldsymbol{\psi}^T \frac{\partial \boldsymbol{\Psi}}{\partial \rho} + \boldsymbol{\kappa}^T \frac{\partial \boldsymbol{\mathcal{H}}}{\partial \rho} \tag{55}$$

Choosing $\boldsymbol{\psi}$ and $\boldsymbol{\kappa}$ such that

$$\boldsymbol{\psi}^T \frac{\partial \boldsymbol{R}}{\partial \boldsymbol{U}} + \boldsymbol{\kappa}^T \frac{\partial \boldsymbol{\mathcal{H}}}{\partial \boldsymbol{U}} + \frac{\partial F}{\partial \boldsymbol{U}} = 0 \tag{56}$$

The parameter $\boldsymbol{\psi}$ and $\boldsymbol{\kappa}$ are known as adjoint vector. Thus, Eq. (55) can be reduced to,

$$\frac{\partial G}{\partial \rho} = \frac{\partial F}{\partial \rho} + \boldsymbol{\psi}^T \frac{\partial \boldsymbol{\Psi}}{\partial \rho} + \boldsymbol{\kappa}^T \frac{\partial \boldsymbol{\mathcal{H}}}{\partial \rho} \tag{57}$$

The above derivation is usually referred to as the discrete adjoint method [47]. The detailed description of this method can be found in Ref. [47].

### 6.3.2 Sensitivity of objective and constraints

Note that element density $x_e$ is chosen as design variable in this section, and sensitivity with respect to Bezier-based representation parameters **X** can be obtained by the chain rule. For periodic boundary conditions, the equilibrium equations can be written as:

$$\boldsymbol{r} = \boldsymbol{R}(\boldsymbol{u}) + (\boldsymbol{A}^T \boldsymbol{\alpha} \boldsymbol{A})\boldsymbol{u} - \boldsymbol{F} - \boldsymbol{A}^T \boldsymbol{\alpha} \boldsymbol{Q} = \boldsymbol{0} \tag{58}$$

Based on design parametrization, the adjoint method described in above section is employed to obtain the sensitivities of the objective and constraint functions, $\boldsymbol{\theta}$, with respect to density $x_e$, given as:

$$\frac{\partial \theta}{\partial x_e} = \frac{\partial \theta(u)}{\partial x_e} + \boldsymbol{\lambda}^T \frac{\partial \boldsymbol{r}}{\partial x_e} \tag{59}$$

The adjoint variable vector $\boldsymbol{\lambda}$ is obtained by solving the following equation:

$$(\boldsymbol{K}^*)^T \boldsymbol{\lambda} = -\left(\frac{\partial \theta(u)}{\partial x_e}\right)^T \tag{60}$$

where

$$\boldsymbol{K}^* = \left(\boldsymbol{K}_t + (\boldsymbol{A}^T \boldsymbol{\alpha} \boldsymbol{A})\right) \tag{61}$$

Note that $\boldsymbol{K}_t$ is the tangent stiffness matrix at the equilibrium state, and superscript $\boldsymbol{T}$ denotes the transpose of the matrix. More details regarding sensitivities of failure constraints for hyperelastic material can be found in Ref. [1].

### 6.4 Initial guess of geometric component distribution

As in previous works [23], random initial guess of geometry component is chosen to initiate optimization. It is feasible to use random initial distribution for linear problem. One weakness of random initialization is that the ends of one geometry component do not always connect with other components. Thus, this non-connectivity issue is undesirable for solving geometry nonlinear problem in that excessive mesh distortion may happen in the gap region during FEM analysis. Hence it is essential to find an initial layout which should be a connected path of geometry components between loads and the boundary conditions. As described by Ref. [24], the value of failure constraint is highly sensitive to a small change of geometric component design variables, and hence some perturbed initial values could lead to unreasonable optimal design. Thus, a reasonable initial value of design variables is of great significance for convergence of optimization progress. However, how to construct an initial connected design is still a tricky problem, especially for nonlinear optimization problem. From our numerical experiments, a density based optimal design can work as a guidance for geometric component initialization. Inspired by this experience, an identification process is proposed in this paper to construct a reasonable initial values of design variables. This identification progress can be divided into two parts. This first part is topology optimization using density-based methods to obtain a coarse layout, which can be used as a design guidance for optimization with geometric component. It is worth to mention that there is no need to reach an ideal 0-1 solution for density based optimization. A coarse layout with a large amount of intermediate densities during optimization progress (i.e. iteration=10) is enough to yield an initial layout. The second part is an identification progress, which can be regarded as an auxiliary optimization problem. The auxiliary optimization problem is formulated as follows:

$$minimize \ \sum(\bar{\bar{X}}_G(\chi_i, \bar{\bar{\rho}}, w) - X_I)^2 \tag{62}$$

where parameters $\chi_i, \bar{\bar{\rho}}, w$ are the design variables of geometric components. $\bar{\bar{X}}_G$ denotes density projection from geometric components. $X_I$ represents objective density from density-based optimization results. Hence, this optimization problem aims to find an optimal initial layout of geometric component by minimizing the difference between geometry projection with desired density distribution from density-based optimal results. Due to the limited parameters needed to be identified, sequential quadratic programming (SQP) method [48] is implemented here to find a local minimum of the cost function. Detailed description of the identification progress will be demonstrated in numerical examples. It is worth to mention that the material layouts do not have distinct difference under three different loading conditions after the few initial optimization iterations (i.e. 10). For simplicity, we apply the density-based optimized material layout under uniaxial tension as initial configuration to the proposed optimization method for all numerical examples.

## 7. Numerical Examples

In this section, we utilize the optimization formulation described above to design metamaterials with different loading conditions. In 2D material design, $100 \times 100$ bilinear square elements with unit length are employed to generate discrete microstructures. Soft material mechanical behavior is described by the Mooney-Rivlin model with $A_{10} = 34, A_{01} = 5.8$ and $K = 2000$. Material softening is assumed to occur at 10% strain under uniaxial tension, which corresponds to failure strain energy value of 1.2. The design symmetry is enforced, which is also described in Ref. [14]. The plane strain assumption is assumed. To initialize the configuration of geometry component, a coarse density results (iteration equals 10) obtained under uniaxial tension boundary condition is plotted in Fig. 13, in which the resulting Bezier skeleton envelopes identified are also presented. For convenience, we use this initial configuration for all numerical examples under different boundary conditions. It is worth to mention that 15 geometry components are selected to represent the density field for all numerical cases, and the number of control points are chosen as five. Note that the MMA algorithm [49] is applied to solve optimization problems.

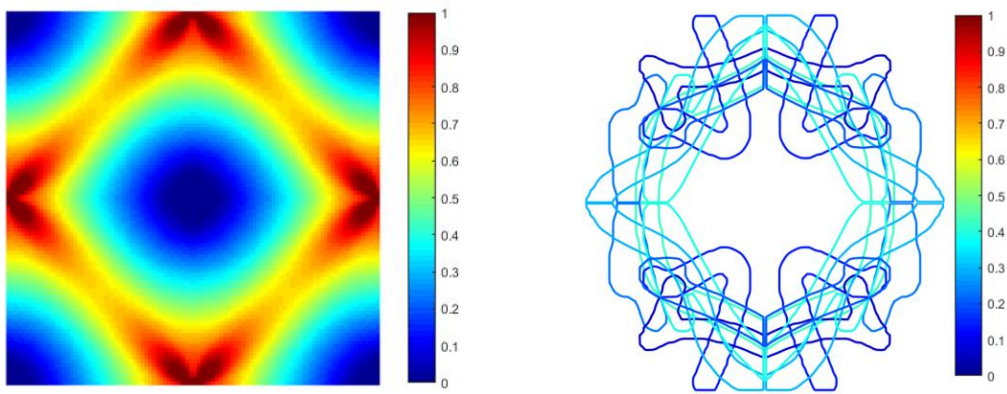

Figure 13. Density-based result based on coarse mesh and the resulting Bezier skeleton envelopes as initial configuration for the proposed optimization method

### 7.1 2D optimized results under uniaxial tension

For material design under uniaxial tension, we design materials with symmetry along both axial directions. To characterize material behavior using the unit cell as presented in Fig. 14, periodic boundary conditions are applied along four sides of the design domain. $\bar{u}$ is a constant displacement difference between right and left edges. The strain energy limit for base material is set to 1.1, which corresponds to 10% strain under uniaxial tension. The goal is to maximize the material stiffness under 30% strain along the horizontal direction without local material failure. The design domain is discretized with $100 \times 100$ quadrilateral elements with element size equals to 1. The loading and boundary conditions together with the design domain are plotted in Fig. 14. The constant displacement difference is set to $\bar{u} = 30$. The width of the Bezier skeleton is set to be $2 < w < 3$, and volume fraction constraint is chosen as 0.3. Note that the value of p-norm for strain energy aggregation is chosen as $p = 10$. In fact, increasing the value of $p$ is better to improve the approximation of the maximum, while a value of $p$ that is too high will result in convergence difficulty during optimization. The optimization begins with the initial configuration demonstrated in Fig. 13. Due to the highly nonlinear nature, the move limit for MMA algorithm need to be small enough to ensure that no gap exists during optimization progress, which will lead to excessive mesh distortion in finite element analysis. The move limit is chosen as $0.01$ after several numerical tests. The optimized material layout is shown in Fig. 15(a). For comparison, the optimized density results without failure constraint is plotted in Fig.15 (b). The optimized Bezier skeleton and enveloping line are presented in Figs. 16 and 17, where only the geometric components with $\bar{\bar{\rho}} > 0.1$ are displayed. Based on optimization results considering failure constraint, only four effective Bezier-based components remain after optimization has been completed, which demonstrates that the optimizer is able to remove redundant geometric components from the initial design. We would like to mention that there is no need to employ higher order Bezier curves, because 15 initial components have enough degrees of freedom to explore the design domain. The $5 \times 5$ lattice structures obtained based on the optimized design are shown in Fig. 18. It is interesting to find that the optimized metamaterial considering failure constraint shares high similarity to the so-called "horseshoe" serpentine design, which is widely used in stretchable electronics as described in Refs. [2, 50-52]. Besides stretchable electronics, these serpentine-shaped structures can be found in many expandable systems made by stiff materials such as cardiovascular stents [53]. The "horseshoe" serpentine structure is presented in Fig. 19. In fact, "horseshoe" microstructures can rotate to accommodate the applied displacement, leaving much smaller intrinsic strain in the base materials compared with the applied strain. The contours of strain energy distribution with a thousand lines are presented in Fig. 20. The maximum local strain energy of optimized results without failure constraint almost reaches 20, while the material stiffness is around 4 times compared to the optimized design considering failure constraint. The strain energy contours on the deformed configuration of the design with failure constraint is presented in Fig. 21.

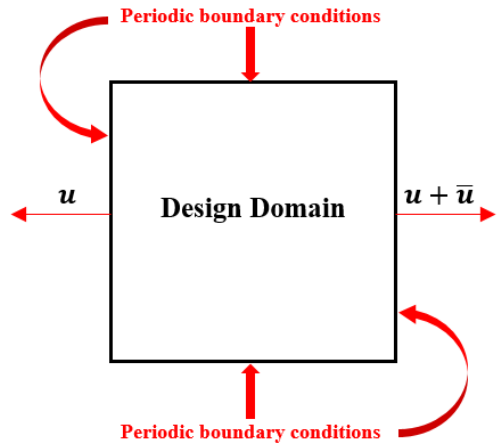

Figure 14. Design domain of a unit cell under uniaxial tension

(a) with failure constraint  (b) without failure constraint

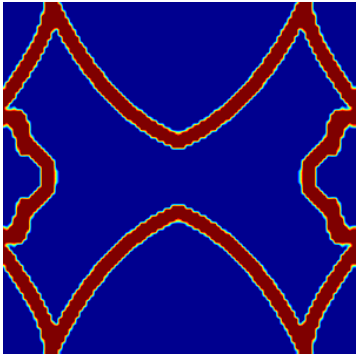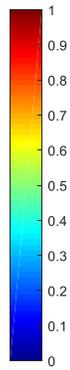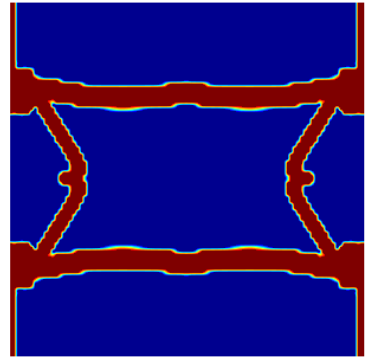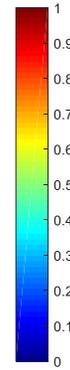

Figure 15. Optimized density results under uniaxial tension (a) with failure constraint and (b) without failure constraint

(b) with failure constraint  (b) without failure constraint

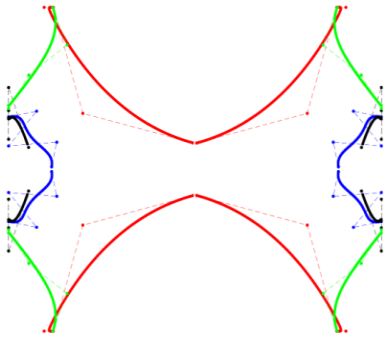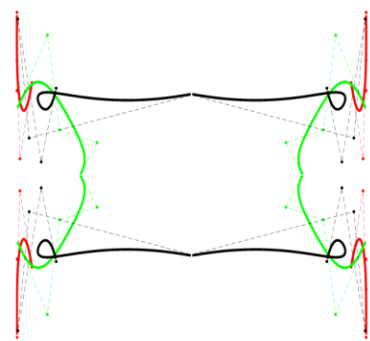

Figure 16. Optimized Bezier skeleton (a) with failure constraint and (b) without failure constraint

(a) with failure constraint  (b) without failure constraint

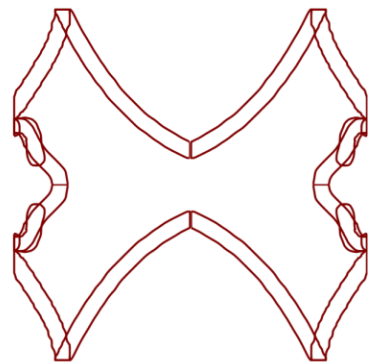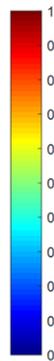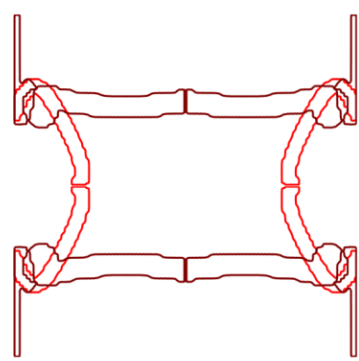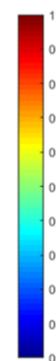

Figure 17. Optimized Bezier skeleton envelopes (a) with failure constraint and (b) without failure constraint

(a) with failure constraint                          (b) without failure constraint

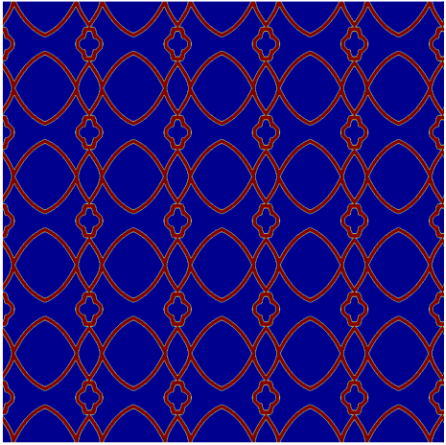 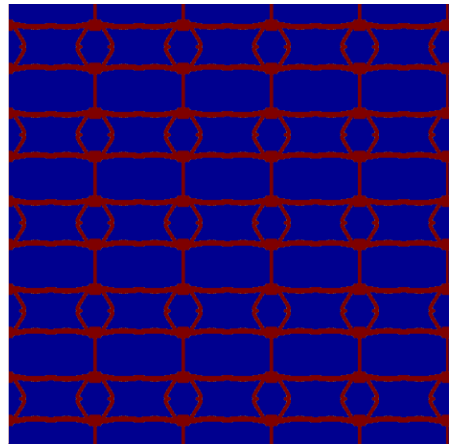

Figure 18. A 5 by 5 lattice structure consisted of the optimized microstructure design (a) with failure constraint and (b) without failure constraint

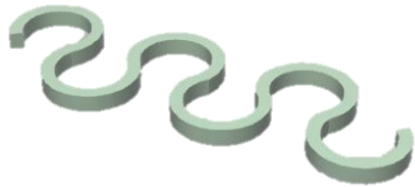

Figure 19. The "horseshoe" serpentine structure

(b) with failure constraint                          (b) without failure constraint

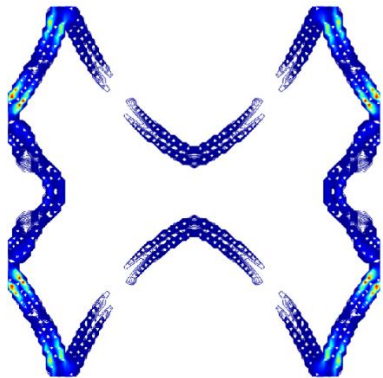 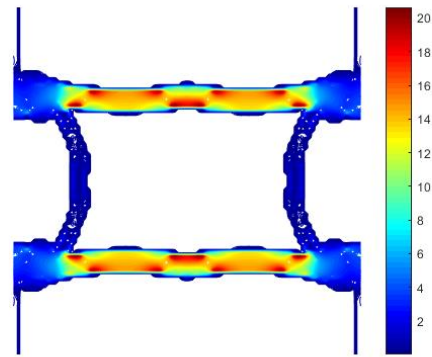

Figure 20. Strain energy distribution for undeformed configuration (a) with failure constraint (b) without failure constraint

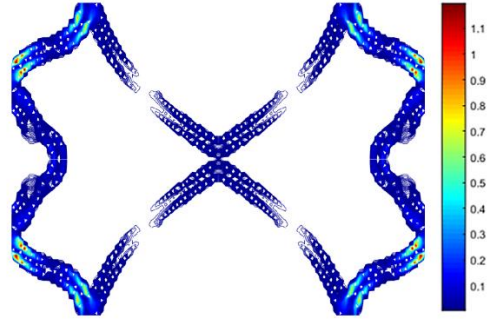

Figure 21. Strain energy distribution contours on deformed configuration

### 7.2 2D optimized results under equal biaxial tension

In this subsection, we apply the proposed optimization formulation to design materials under equal biaxial tension, see Fig. 22. We aim at designing high stiffness materials subjected to 30% strain along both longitudinal and transverse directions without local failure. The dimension and material properties are the same as previous numerical example, where $100 \times 100$ quadrilateral elements with element size equals to 1 are used to discretize the design domain. The same optimization parameters are applied in this numerical example. The maximum allowed volume is 30% of the design domain volume. The optimized microstructure considering failure constraint is presented in Fig. 23(a), while the result without constraint is shown in Fig. 23(b). Figures 24 and 25 show the optimized Bezier skeletons and enveloping lines. The optimized result show "horseshoe" structures along both directions as shown in Fig. 23. Note that only the geometric components with $\bar{\bar{\rho}} > 0.1$ are displayed. For design without failure constraint, the optimized microstructure shares much similarity with the honeycomb structure, see Fig. 23(b). The perspective of the full periodic microstructure is demonstrated in Fig. 26. The strain energy contours are found in Fig. 27, where deformation at 30% strain along both directions is plotted in Fig. 28.

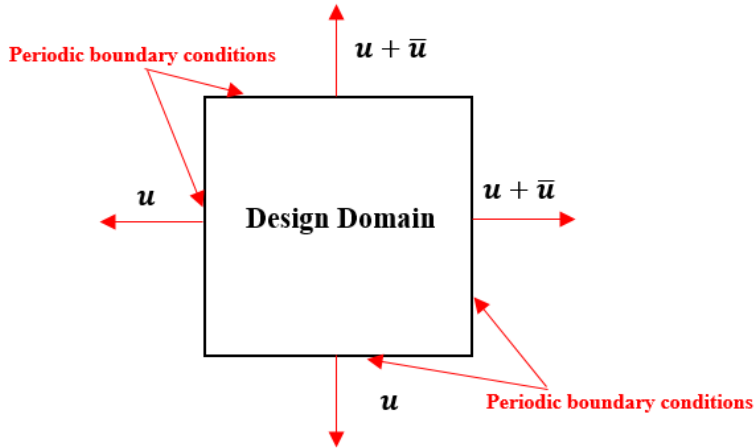

Figure 22. Design domain of a unit cell under equal biaxial tension

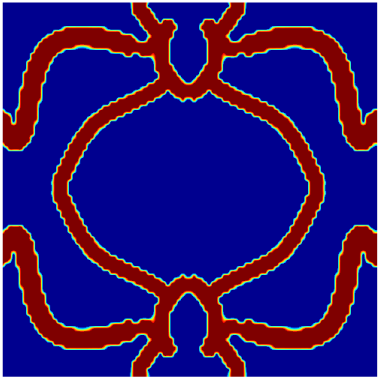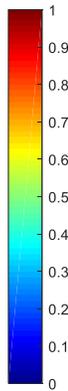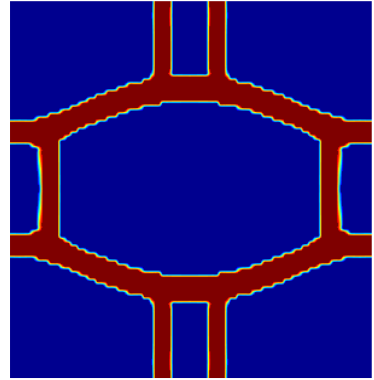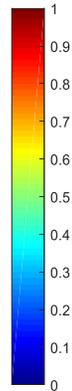

Figure 23. Optimized density results under equal biaxial tension (a) with failure constraint and (b) without failure constraint

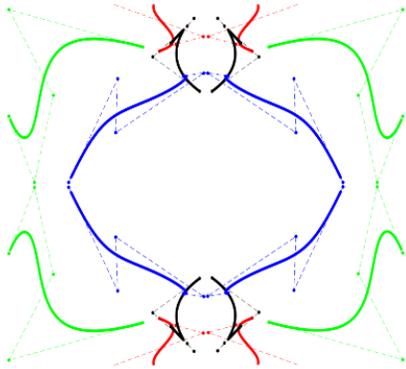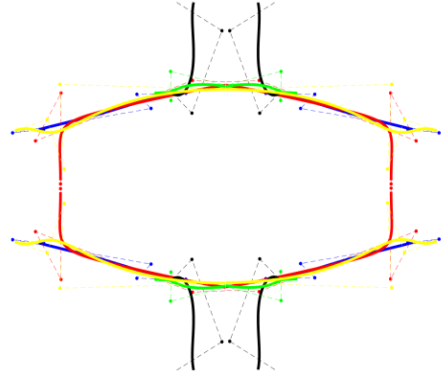

Figure 24. Optimized Bezier skeleton (a) with failure constraint and (b) without failure constraint

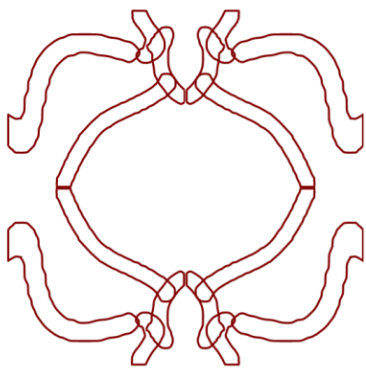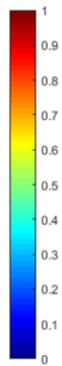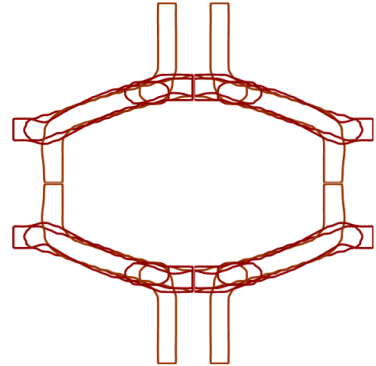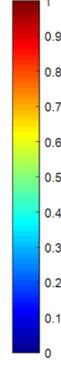

Figure 25. Optimized Bezier skeleton envelopes (a) with failure constraint and (b) without failure constraint

(a) with failure constraint (b) without failure constraint

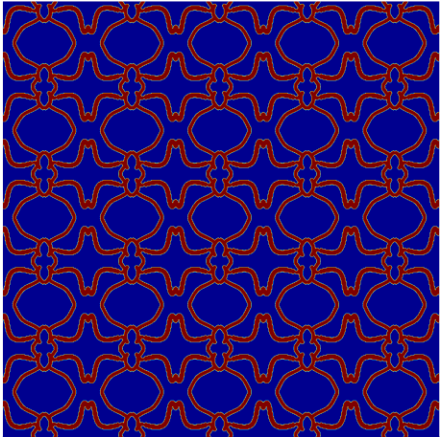 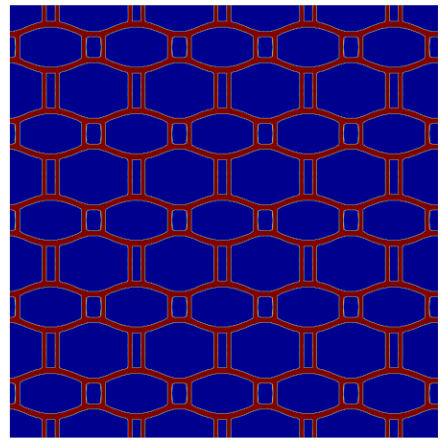

Figure 26. A 5 by 5 lattice structure consisted of the optimized microstructure (a) with failure constraint and (b) without failure constraint

(a) with failure constraint (b) without failure constraint

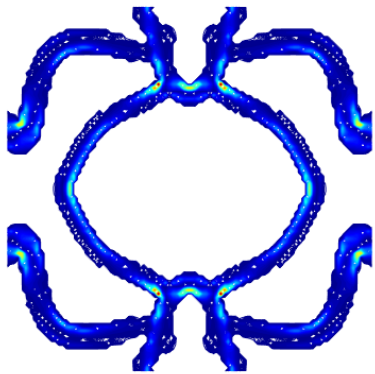 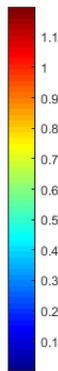 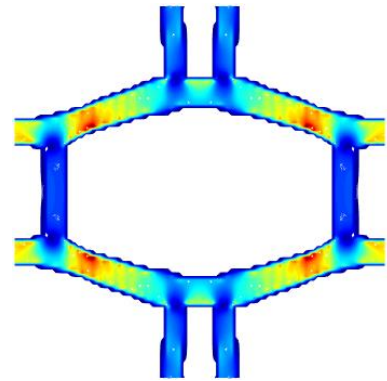 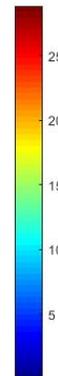

Figure 27. Strain energy distribution on undeformed configuration (a) with failure constraint and (b) without failure constraint

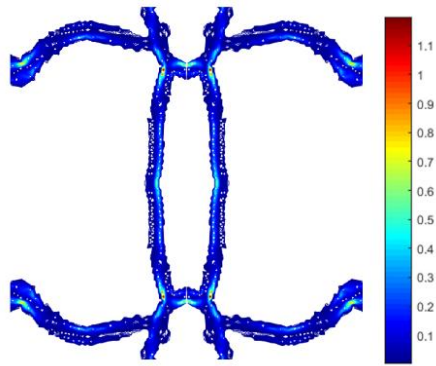

Figure 28. Strain energy distribution contours on deformed configuration

### 7.3 2D optimized results under pure shear

In this section, for the stretchable design of lattice structures, a single unit cell is isolated in the repeating pattern as shown in Fig. 29. The material layout in this unit cell is optimized to create high stiffness metamaterial, which can subjected to 30% shear strain without local material failure. Periodic boundary conditions are implemented to simulate material macro-behavior. The boundary conditions applied on the unit cell is presented in Fig. 29. The design domain is discretized by 4-node $100 \times 100$ elements with unit length. Using the approach described before, the optimization progress begins with the initial configuration plotted in Fig. 13, where the volume fraction is set to be 30% in this numerical case. In the optimized design, the near circular configuration is presented, while the design without failure constraint is demonstrated in Fig. 30(b) for comparison. The optimized material layout in the 5 by 5 lattice structure are plotted in Fig. 16. To easily compare the strain energy distribution for two different optimized configurations, strain energy contours with a thousand contour lines are plotted in Fig. 34. Note that the local strain energy almost reaches 16 for the design not considering failure constraint. The deformation of optimized microstructure under 30% shear strain is shown in Fig. 35.

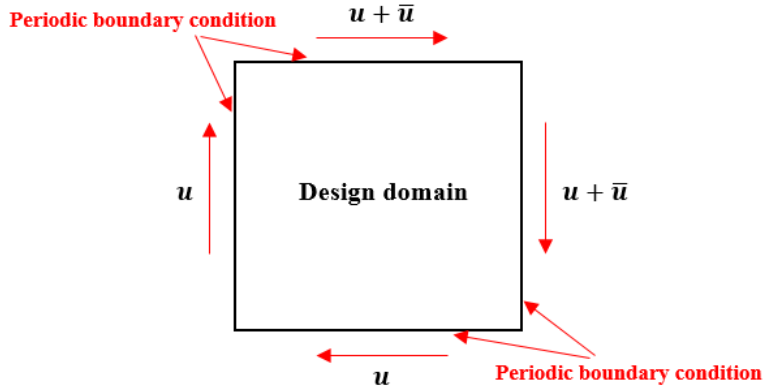

Figure 29. Design domain of a unit cell under pure shear

(a) with failure constraint    (b) without failure constraint

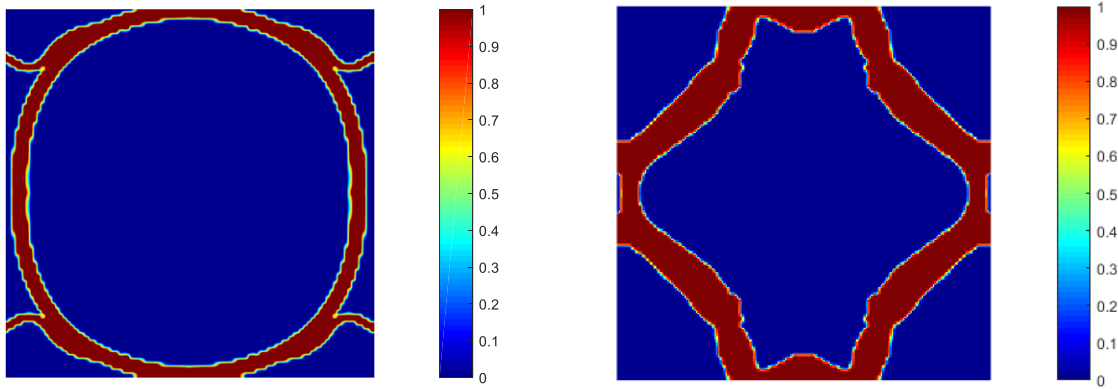

Figure 30. Optimized density results under equal biaxial tension (a) with failure constraint (b) without failure constraint

(a) with failure constraint    (b) without failure constraint

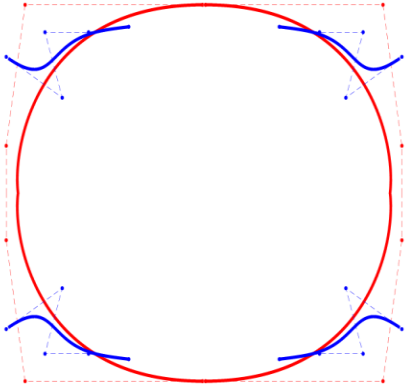 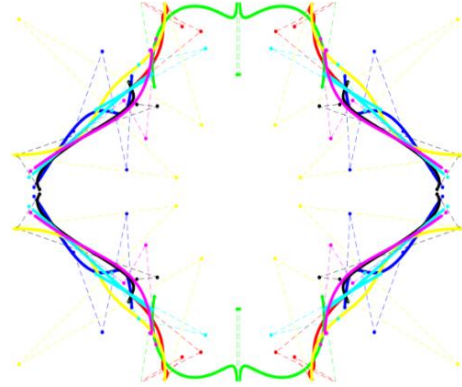

Figure 31. Optimized Bezier skeleton (a) with failure constraint (b) without failure constraint

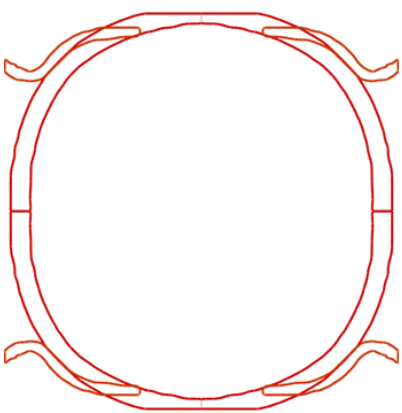 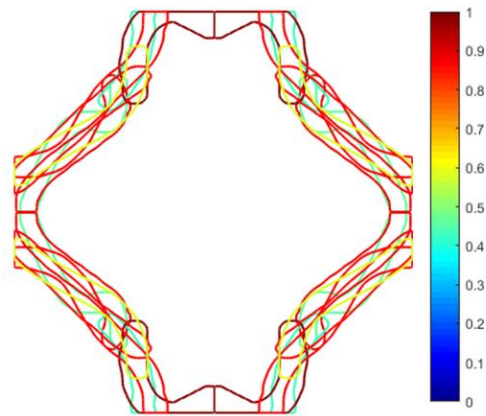

Figure 32. Enveloping line of optimized Bezier skeleton (a) with failure constraint (b) without failure constraint

(b) with failure constraint            (b) without failure constraint

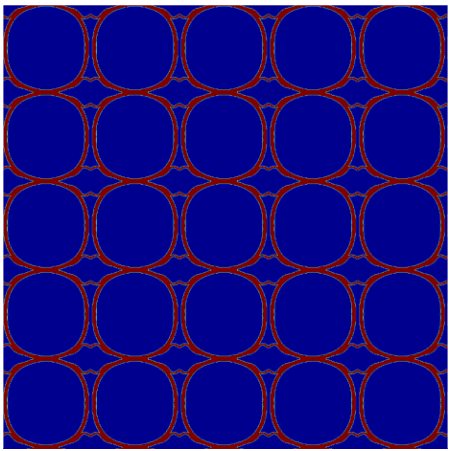 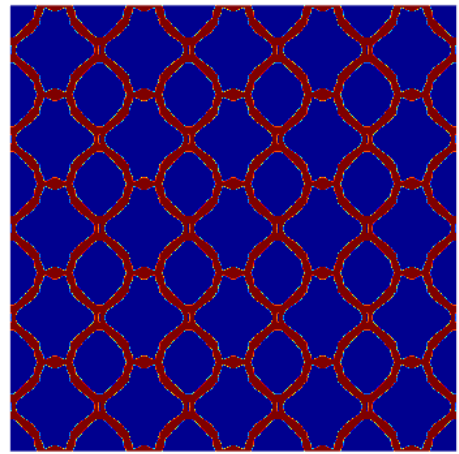

Figure 33. 5 by 5 lattice structure (a) with failure constraint (b) without failure constraint

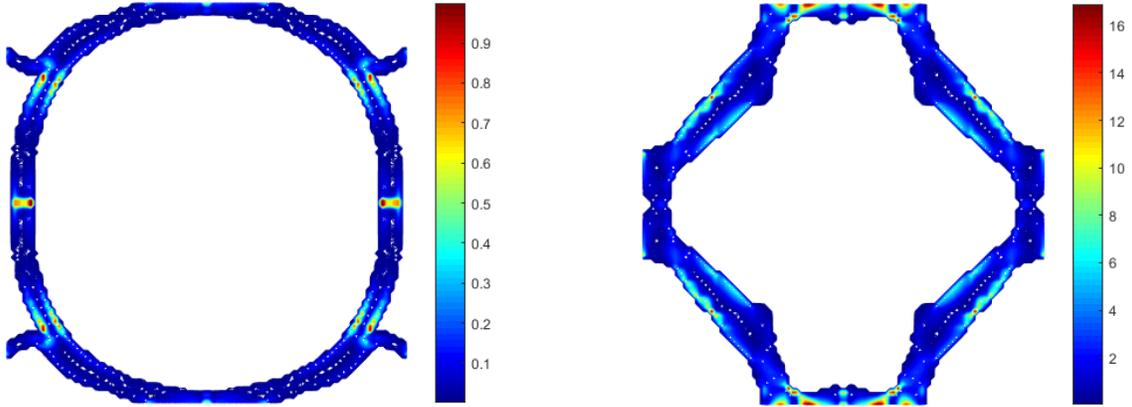

Figure 34. Strain energy distribution contours on undeformed configuration (a) with failure constraint (b) without failure constraint

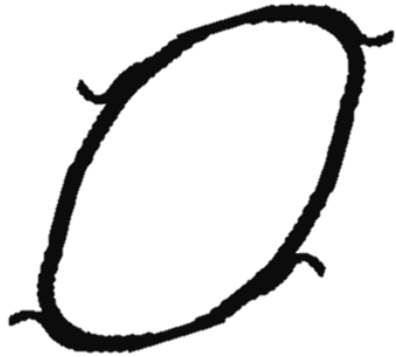

Figure 35. Deformation of optimized microstructure under 30% shear strain

## 8. Conclusion

In this paper, a Bezier skeleton explicit density (BSED) representation algorithm is proposed for the topology optimization of stretchable metamaterial. Material failure is measured by strain energy and p-norm formulation are utilized. A Heaviside function is applied to create a mapping from geometry skeleton to mesh grids, where the skeleton is described by the Bezier curves. This density representation method successfully inherits the main advantages of density-based topology optimization. Sensitivities of the objectives and constraints with respect to control parameters can be readily derived by using the chain rule. Standard nonlinear programming algorithms are applied in this algorithm. The initial configuration for the proposed method is obtained by performing optimization using standard density-based method on coarse mesh for a few iterations, which would lead to a well-connected layout. From the numerical examples, redundant geometry members are able to be removed, and the thickness of skeleton can be easily controlled by parameters of Heaviside function. Due to the powerful curve fitting ability, using Bezier curve to represent density field can explore design space effectively compared to bar-like structures, and generate manufacturing friendly structures without any intricate small features in optimal design. Furthermore, this density representation method is mesh independent and the design variables are reduced significantly so that the optimization problem can be solved efficiently using regular optimization algorithm. From the numerical results, the optimized material layout shares high similarity to the "horseshoe" structures, which are widely found in soft electronics design. Therefore, the method proposed in this paper shows great potential and opens the door for designing manufacturable microstructures to achieve extreme stretchable materials that can be utilized for applications such as stretchable electronics and soft robotics.

# Reference

[1] H. Deng, L. Cheng, A. C. J. S. To, and M. Optimization, "Distortion energy-based topology optimization design of hyperelastic materials," pp. 1-19, 2018.

[2] T. Widlund, S. Yang, Y.-Y. Hsu, N. J. I. J. o. S. Lu, and Structures, "Stretchability and compliance of freestanding serpentine-shaped ribbons," vol. 51, no. 23-24, pp. 4026-4037, 2014.

[3] Y. Jiang and Q. Wang, "Highly-stretchable 3D-architected mechanical metamaterials," *Scientific reports,* vol. 6, p. 34147, 2016.

[4] O. Sigmund, S. J. J. o. t. M. Torquato, and P. o. Solids, "Design of materials with extreme thermal expansion using a three-phase topology optimization method," vol. 45, no. 6, pp. 1037-1067, 1997.

[5] D. C. Dobson and S. J. J. S. J. o. A. M. Cox, "Maximizing band gaps in two-dimensional photonic crystals," vol. 59, no. 6, pp. 2108-2120, 1999.

[6] O. Sigmund, J. S. J. P. T. o. t. R. S. o. L. A. M. Jensen, Physical, and E. Sciences, "Systematic design of phononic band–gap materials and structures by topology optimization," vol. 361, no. 1806, pp. 1001-1019, 2003.

[7] F. Wang, O. Sigmund, and J. S. Jensen, "Design of materials with prescribed nonlinear properties," *Journal of the Mechanics and Physics of Solids,* vol. 69, pp. 156-174, 2014.

[8] E. Andreassen, B. S. Lazarov, and O. Sigmund, "Design of manufacturable 3D extremal elastic microstructure," *Mechanics of Materials,* vol. 69, no. 1, pp. 1-10, 2014.

[9] A. Clausen, F. Wang, J. S. Jensen, O. Sigmund, and J. A. Lewis, "Topology optimized architectures with programmable Poisson's ratio over large deformations," *Advanced Materials,* vol. 27, no. 37, pp. 5523-5527, 2015.

[10] C. R. Thomsen, F. Wang, and O. Sigmund, "Buckling strength topology optimization of 2D periodic materials based on linearized bifurcation analysis," *Computer Methods in Applied Mechanics and Engineering,* vol. 339, pp. 115-136, 2018.

[11] J.-C. Michel, H. Moulinec, P. J. C. m. i. a. m. Suquet, and engineering, "Effective properties of composite materials with periodic microstructure: a computational approach," vol. 172, no. 1-4, pp. 109-143, 1999.

[12] P. P. J. J. o. t. M. Castañeda and P. o. Solids, "Exact second-order estimates for the effective mechanical properties of nonlinear composite materials," vol. 44, no. 6, pp. 827-862, 1996.

[13] N.-H. Kim, *Introduction to nonlinear finite element analysis*. Springer Science & Business Media, 2014.

[14] F. Wang, O. Sigmund, J. S. J. J. o. t. M. Jensen, and P. o. Solids, "Design of materials with prescribed nonlinear properties," vol. 69, pp. 156-174, 2014.

[15] M. P. Bendsøe, A. Ben-Tal, and J. Zowe, "Optimization methods for truss geometry and topology design," *Structural optimization,* vol. 7, no. 3, pp. 141-159, 1994.

[16] T. Hagishita and M. Ohsaki, "Topology optimization of trusses by growing ground structure method," *Structural and Multidisciplinary Optimization,* vol. 37, no. 4, pp. 377-393, 2009.

[17] A. Asadpoure, J. K. Guest, and L. Valdevit, "Incorporating fabrication cost into topology optimization of discrete structures and lattices," *Structural and Multidisciplinary Optimization,* vol. 51, no. 2, pp. 385-396, 2015.

[18] A. J. Torii, R. H. Lopez, and L. F. Miguel, "Design complexity control in truss optimization," *Structural and Multidisciplinary Optimization,* vol. 54, no. 2, pp. 289-299, 2016.

[19] X. Guo, W. Zhang, and W. Zhong, "Doing topology optimization explicitly and geometrically—a new moving morphable components based framework," *Journal of Applied Mechanics,* vol. 81, no. 8, p. 081009, 2014.

[20] W. Zhang, D. Li, J. Zhou, Z. Du, B. Li, and X. Guo, "A Moving Morphable Void (MMV)-based explicit approach for topology optimization considering stress constraints," *Computer Methods in Applied Mechanics and Engineering,* vol. 334, pp. 381-413, 2018.
[21] W. Zhang *et al.*, "Topology optimization with multiple materials via moving morphable component (MMC) method," *International Journal for Numerical Methods in Engineering,* vol. 113, no. 11, pp. 1653-1675, 2018.
[22] W. Zhang *et al.*, "Explicit three dimensional topology optimization via Moving Morphable Void (MMV) approach," *Computer Methods in Applied Mechanics and Engineering,* vol. 322, pp. 590-614, 2017.
[23] J. Norato, B. Bell, and D. Tortorelli, "A geometry projection method for continuum-based topology optimization with discrete elements," *Computer Methods in Applied Mechanics and Engineering,* vol. 293, pp. 306-327, 2015.
[24] S. Zhang, A. L. Gain, and J. A. Norato, "Stress-based topology optimization with discrete geometric components," *Computer Methods in Applied Mechanics and Engineering,* vol. 325, pp. 1-21, 2017.
[25] S. Watts and D. A. Tortorelli, "A geometric projection method for designing three‐dimensional open lattices with inverse homogenization," *International Journal for Numerical Methods in Engineering,* vol. 112, no. 11, pp. 1564-1588, 2017.
[26] D. A. White, M. L. Stowell, D. A. J. S. Tortorelli, and M. Optimization, "Toplogical optimization of structures using Fourier representations," vol. 58, no. 3, pp. 1205-1220, 2018.
[27] M. Mooney, "A theory of large elastic deformation," *Journal of applied physics,* vol. 11, no. 9, pp. 582-592, 1940.
[28] S. Zhang, A. L. Gain, and J. A. Norato, "A geometry projection method for the topology optimization of curved plate structures with placement bounds," *International Journal for Numerical Methods in Engineering,* vol. 114, no. 2, pp. 128-146, 2018.
[29] X.-D. Chen, J.-H. Yong, G. Wang, J.-C. Paul, and G. J. C.-A. D. Xu, "Computing the minimum distance between a point and a NURBS curve," vol. 40, no. 10-11, pp. 1051-1054, 2008.
[30] A. Matsuda and N. J. P. E. Kawasaki, "Computational Homogenization Analysis Applied to Hyperelasticity for Porous Polymers," vol. 34, pp. 706-711, 2012.
[31] Y. Shinoda and A. J. P. E. Matsuda, "Homogenization analysis of porous polymer considering microscopic structure," vol. 60, pp. 343-348, 2013.
[32] Q. Yang and F. J. F. o. M. E. i. C. Xu, "Numerical modeling of nonlinear deformation of polymer composites based on hyperelastic constitutive law," vol. 4, no. 3, pp. 284-288, 2009.
[33] Q. Gu, M. Barbato, and J. P. J. J. o. E. M. Conte, "Handling of constraints in finite-element response sensitivity analysis," vol. 135, no. 12, pp. 1427-1438, 2009.
[34] Q. J. I. J. o. S. Li and Structures, "Strain energy density failure criterion," vol. 38, no. 38-39, pp. 6997-7013, 2001.
[35] M. Kipp, G. J. I. J. o. S. Sih, and Structures, "The strain energy density failure criterion applied to notched elastic solids," vol. 11, no. 2, pp. 153-173, 1975.
[36] P. Trapper and K. Volokh, "Elasticity with energy limiters for modeling dynamic failure propagation," *International Journal of Solids and Structures,* vol. 47, no. 25-26, pp. 3389-3396, 2010.
[37] K. Volokh and H. Gao, "On the modified virtual internal bond method," *Journal of Applied Mechanics,* vol. 72, no. 6, pp. 969-971, 2005.
[38] K. Volokh, "Hyperelasticity with softening for modeling materials failure," *Journal of the Mechanics and Physics of Solids,* vol. 55, no. 10, pp. 2237-2264, 2007.
[39] P. Trapper and K. Volokh, "Modeling dynamic failure in rubber," *International Journal of Fracture,* vol. 162, no. 1-2, pp. 245-253, 2010.
[40] K. Volokh and P. Trapper, "Fracture toughness from the standpoint of softening hyperelasticity," *Journal of the Mechanics and Physics of Solids,* vol. 56, no. 7, pp. 2459-2472, 2008.

[41] K. Y. Volokh, "Multiscale modeling of material failure: From atomic bonds to elasticity with energy limiters," *International Journal for Multiscale Computational Engineering,* vol. 6, no. 5, 2008.

[42] D. Rittel, Z. Wang, and M. Merzer, "Adiabatic shear failure and dynamic stored energy of cold work," *Physical review letters,* vol. 96, no. 7, p. 075502, 2006.

[43] K. Volokh, "Comparison of biomechanical failure criteria for abdominal aortic aneurysm," *Journal of biomechanics,* vol. 43, no. 10, pp. 2032-2034, 2010.

[44] K. J. J. o. t. m. b. o. b. m. Volokh, "Modeling failure of soft anisotropic materials with application to arteries," vol. 4, no. 8, pp. 1582-1594, 2011.

[45] F. Wang, B. S. Lazarov, O. Sigmund, and J. S. Jensen, "Interpolation scheme for fictitious domain techniques and topology optimization of finite strain elastic problems," *Computer Methods in Applied Mechanics and Engineering,* vol. 276, pp. 453-472, 2014.

[46] E. Holmberg, B. Torstenfelt, A. J. S. Klarbring, and M. Optimization, "Stress constrained topology optimization," vol. 48, no. 1, pp. 33-47, 2013.

[47] K. C. Giannakoglou and D. I. Papadimitriou, "Adjoint methods for shape optimization," in *Optimization and computational fluid dynamics*: Springer, 2008, pp. 79-108.

[48] J.-F. Bonnans, J. C. Gilbert, C. Lemaréchal, and C. A. Sagastizábal, *Numerical optimization: theoretical and practical aspects*. Springer Science & Business Media, 2006.

[49] K. J. U. h. p. k. s. k. m. p. Svanberg, "MMA and GCMMA-two methods for nonlinear optimization," 2007.

[50] Z. Fan *et al.*, "A finite deformation model of planar serpentine interconnects for stretchable electronics," vol. 91, pp. 46-54, 2016.

[51] Y. Zhang *et al.*, "Mechanics of ultra-stretchable self-similar serpentine interconnects," vol. 61, no. 20, pp. 7816-7827, 2013.

[52] Q. Ma *et al.*, "A nonlinear mechanics model of bio-inspired hierarchical lattice materials consisting of horseshoe microstructures," vol. 90, pp. 179-202, 2016.

[53] R. Beyar *et al.*, "Multicenter Pilot Study of a Serpentine Balloon‐Expandable Stent (beStentTM): Acute Angiographic and Clinical Results," vol. 10, no. 4, pp. 277-286, 1997.